\documentclass[a4paper]{jpconf}
\usepackage{graphicx, xcolor, latexsym}
\usepackage{amsmath, amsfonts, amssymb}
\usepackage[numbers,square,sort&compress]{natbib}
\usepackage{hyperref}

\newcommand{\ud}{\mathrm{d}}

\begin{document}


\title{Generalized Master equation approach to mesoscopic time-dependent transport}

\author{Kristinn Torfason$^{1,2}$, Andrei Manolescu$^1$, Valeriu Molodoveanu$^3$, 
Vidar Gudmundsson$^2$}
\address{$^1$ School of Science and Engineering, Reykjavik University,
Menntavegur 1, IS-101 Reykjavik, Iceland}
\address{$^2$ Science Institute, University of Iceland, Dunhaga 3,
IS-107 Reykjavik, Iceland}
\address{$^3$ National Institute of Materials Physics, P. O. Box MG-7, 
Bucharest-Magurele, Romania}

\ead{manoles@ru.is}


\begin{abstract}
We use a generalized Master equation (GME) formalism to describe the
non-equilibrium time-dependent transport through a short quantum wire
connected to semi-infinite biased leads. The contact strength between
the leads and the wire are modulated by out-of-phase time-dependent
functions which simulate a turnstile device.  One lead is fixed at one
end of the sample whereas the other lead has a variable placement.
The system is described by a lattice model.  We find that the currents
in both leads depend on the placement of the second lead.  In the rather
small bias regime we obtain transient currents flowing against the bias
for short time intervals.  The GME is solved numerically in small time
steps without resorting to the traditional Markov and rotating wave
approximations.  The Coulomb interaction between the electrons in the
sample is included via the exact diagonalization method.

\end{abstract}







\section{Introduction}

The theoretical description of time-dependent transport in
semiconductor nanostructures has received considerable attention
in the last few years. Non-equilibrium Greens' function
techniques and density-functional methods were developed for transient current
calculations in interacting and non-interacting structures (see
\cite{Kurth1,Stefanucci,Myohanen1,Myohanen2} and references therein).
These methods were employed to study the response of a mesoscopic sample
to a time-dependent (possibly pulsed) voltage applied on the leads and/or
to check the expected crossover to a steady-state.

A strong motivation behind these studies is the need to model and predict
the transient response of open and interacting nanodevices subjected
to time-dependent signals. Since in real systems both the contacts and
sample geometry as well as the charging and correlation effects are
important, the numerical implementation of the various formal methods
requires extensive and costly computational work.

Recently we reported transport calculations for a two-dimensional
parabolic quantum wire in the turnstile setup \cite{NJP}, where the
Coulomb interaction between electrons was neglected.  The wire 
was in contact with external leads, seen as particle reservoirs, and
the contact regions were described by a phenomenological ansatz. We
provided a careful analysis of the electronic propagation on the edge
states that develop in the presence of a strong perpendicular
magnetic field.  Let us remind here that the turnstile setup was
experimentally realized by Kouwenhoven {\it et al.} \cite{TSP}.
It essentially involves a time-dependent modulation (pumping) of the
tunneling barriers between the finite sample and drain and source leads,
respectively. During the first half of the pumping cycle the
system opens only to the source lead whereas during the second half of
the cycle the drain contact opens. At certain parameters an integer
number of electrons is transferred across the sample in a complete
cycle. Similar transient current measurements were performed in a
pump-and-probe configuration \cite{Tarucha,Lai,Naser}. More complex
turnstile pumps have been studied by numerical simulations, like
one-dimensional arrays of junctions \cite{Mizugaki} or two-dimensional
multidot systems \cite{Ikeda}. The turnstile regime differs from the
adiabatic quantum pumping where charge is transferred along a sample
even in the absence of a bias.

In this paper we describe the turnstile regime of a
one-dimensional (1D) quantum wire (``the sample'').  The effect of the
electron-electron interaction is included in the sample via the 
exact diagonalization method while the time-dependent transport is
performed within the generalized Master equation (GME) formalism as it
is described in Ref.\ \cite{PRBC}.  The implemented GME formalism can be
used to describe both the initial transient regime immediately after the
coupling of the leads to the sample and the evolution towards a steady
state achieved in the long time limit.  
To the best of our knowledge these are the first numerical simulations
of electronic transport through a Coulomb interacting and spatially
extended quantum turnstile.
We discuss for the first time the effect of  contacts' location on the
transient currents. More precisely, we show that if the drain lead is
attached to different regions of the quantum wire the currents in both
leads are considerably affected.

The paper is organized as follows: The model and the methodology are
described in Section\ \ref{sec:model},
the numerical results are presented in Section\ \ref{sec:results},
and the conclusions in Section\ \ref{sec:conclusion}.


\section{The physical model and the GME}\label{sec:model}



\subsection{Setup}

\begin{figure}
    \centering
      \includegraphics{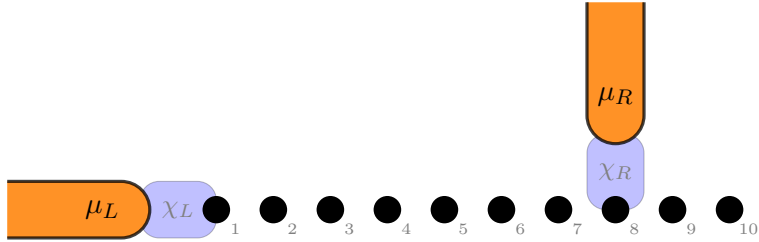}
\caption{A sketch of the system under study. A 1D lattice with 10 sites 
(``the sample'') is connected to two semi-infinite leads via tunneling.
The left lead is connected to the left end of lattice, while the position
of the right lead can be changed.  The contacts are modulated in time with
the functions $\chi_{L,R}$ given in Eq.\ \ref{eq:turnstile}.
}
    \label{fig:GME-system}
  \end{figure}

The physical system consists in a sample connected to two leads acting
as particle reservoirs.  We shall adopt a tight-binding description of
the system: the sample is a 10-site quantum wire and the leads are 1D
and semi-infinite.  A sketch is given in Fig.\ \ref{fig:GME-system}.
The left lead (or the source, marked as $L$) is contacted at one end
of the sample and the right lead (or the drain, marked as $R$) may be
contacted on any other site.  The Hamiltonian of the coupled and electrically
biased system reads as
  \begin{equation}
    H(t) = \sum_{\ell} H_{\ell} + H_S + H_T(t) = H_0 + H_T(t)\, ,
  \end{equation}
where $H_S$ is the Hamiltonian of the isolated sample, including the
electron-electron interaction, and $H_{\ell}$, with  $\{\ell\} = (L,
R)$, corresponds to the left and the right leads.  $H_T$ describes the
time-dependent coupling between the single-particle basis states $\{
|\phi_n\rangle \}$ of the isolated sample and the states $\{ {\psi}_{q\ell}
\}$ of the leads:
\begin{equation}\label{Htunnel}
H_T(t)=\sum_{n}\sum_{\ell}\int \ud q\:\chi_{\ell}(t)(T^{\ell}_{qn}c^{\dagger}_{q\ell}d_n + h.c.) \,.
\end{equation}
The function $\chi_{\ell}(t)$ describes the time-dependent switching
of the sample-lead contacts, while $d^{\dagger}_n$ and $ c_{q\ell}$
create/annihilate electrons in the corresponding single-particle
states of the sample or leads, respectively. The coupling coefficient
$T^{\ell}_{qn}=V_0{\psi}^{*}_{q\ell}(0)\phi_n(i_\ell)$ involves the two
eigenfunctions evaluated at the contact sites $(0,i_\ell)$, $0$ being the
site of the lead $\ell$ and $i_\ell$ the site in the sample \cite{PRBC}.
In our present calculations we keep the left lead connected to the site
$i_L=1$, while the position of the right lead is $i_R=10$ or $i_R=3$.
The parameter $V_0$ plays the role of a coupling constant between the 
sample and the leads.


\subsection{GME}

Following the Nakajima-Zwanzig technique \cite{PhysRevB.77.195416}
we define the reduced density operator (RDO), $\rho(t)$, by tracing
out the degrees of freedom of the environment, the leads in our case,
over the statistical operator of the entire system, $W(t)$:
  \begin{equation}\label{eq:rdo}
    \rho(t) = \tr_L\tr_R W(t)\, ,\quad \rho(0) = \rho_S\, .
  \end{equation}
The initial condition corresponds to a decoupled sample and leads when the RDO is
just the statistical operator of the isolated sample $\rho_S$.  For a sufficiently
weak coupling strength ($V_0$) one obtains
the non-Markovian integro-differential Master equation for the RDO:
\begin{eqnarray}\nonumber
{\dot\rho}(t)=-\frac{i}{\hbar}[H_S,\rho(t)]-\frac{1}{\hbar^2}\sum_{\ell}\int \ud q\:\chi_{\ell}(t)
([{\cal T}_{q\ell},\Omega_{q\ell}(t)]+h.c.) \, ,
\end{eqnarray}
where the operators $\Omega_{q\ell}$ and $\Pi_{q\ell}$ are defined as
\begin{eqnarray}\nonumber
&&\Omega_{q\ell}(t)=e^{-itH_S} \int_{0}^t\,\ud s\:\chi_{\ell}(s)\Pi_{q\ell}(s)e^{i(s-t)
\varepsilon_{q\ell}}e^{itH_S} \, ,\\\nonumber
&&\Pi_{q\ell}(s)=e^{isH_S}\left ({\cal T}_{q\ell}^{\dagger}\rho(s)(1-f_\ell)-\rho(s)
{\cal T}_{q\ell}^{\dagger}f_\ell\right )e^{-isH_S} \, ,
\end{eqnarray}
and $f_\ell$ is the Fermi function of the lead $\ell$. The operators ${\cal T}_{q\ell}$ and
${\cal T}_{q\ell}^{\dagger}$ describe the 'transitions' between two many-electron states (MES)
$|\alpha\rangle$ and $|\beta\rangle$ when one electron enters the sample or leaves it:
\begin{equation}\label{Tmunuint}
\left( {\cal T}_{q\ell}\right)_{\alpha\beta}=\sum_n 
T^\ell_{qn}\langle\alpha|d^{\dagger}_n|\beta\rangle \,.
\end{equation}
The GME is solved numerically by calculating the matrix elements of the
RDO in the basis of the interacting MES, in small time steps, following a
Crank-Nicolson algorithm.  Using the RDO we obtain the time dependent
charge and currents in the system. More details can be found in Ref.\
\cite{PRBC}.


\subsection{Coulomb interaction}

We will ignore the Coulomb effects in the leads, where we assume a high
concentration of electrons and thus strong screening and fast particle
rearrangements.  The Coulomb electron-electron interaction is considered
in detail only in the sample, where Coulomb blocking effects may occur.
We calculate the MES in the sample following the exact diagonalization
method, i.\ e.\ without any mean field approximation.  The interacting
MES are calculated in the Fock space as superpositions of non-interacting
MES derived from Slater determinants \cite{PRBC}.  Since the sample is
open the number of electrons is not fixed, but the Coulomb interaction
conserves the number of electrons.  With 10 lattice sites we obtain 10
single-particle eigenstates and thus $2^{10}=1024$ elements in the Fock
space spanned by the occupation numbers.  The Coulomb effects are measured
by the ratio of a characteristic Coulomb energy $U_C=e^2/(\kappa a)$
and the hopping energy $t_s=\hbar^2/(2m_{eff}a^2)$. Here $a$ denotes the
inter-site distance (the lattice constant of the discretized system),
while $\kappa$ and $m_{eff}$ are material parameters, the dielectric
constant and the electron effective mass, respectively. In our calculations
the {\em relative} strength of the Coulomb interaction, $u_C=U_C/t_s$. is
considered a free parameter.  We will use $u_C=0.5$.  For a material like
GaAs this value would correspond to a sample length of $9a \approx 22$ nm.
This length may not look very realistic, but the choice of the parameters
was determined by the computational time spent in solving the GME which grows
very fast with the number of MES.

The chemical potentials in the leads create a bias window
$\Delta \mu = \mu_L-\mu_R$. The MES of the sample which participate in the transport
correspond to chemical potentials $\mu_N^{(i)}:={\cal E}_N^{(i)}-{\cal
E}_{N-1}^{(0)}$ situated within the bias window, or maybe only slightly
outside, depending on the sample-leads coupling constant $V_0$
\cite{PRBC}.  Here ${\cal E}_{N}^{(i)}$ is an energy of the sample
spectrum containing $N$ particles, $i=0$ indicating to the ground state
and $i>0$ the excited states.  A diagram of the chemical potentials is shown
in Fig.\ \ref{fig:mu-diag}.


\subsection{Time-dependent switching}

The switching-functions in Eq.\ \ref{Htunnel} act on the contact regions 
shaded \textcolor{blue!85}{blue} in Fig.\ \ref{fig:GME-system} and are used 
to mimic potential barriers with time dependent height.  In the present study
they are made by combining two quasi Fermi functions that are shifted relatively
to each other,
  \begin{equation}
    \chi_\ell(t) = 1 - \frac{1}{e^{t-\gamma_s^\ell-\delta} + 1}
                 - \frac{1}{e^{-(t - \gamma_s^\ell) + (T_p^\ell+\delta)} + 1} \, ,
\qquad t \in [0,\, 2 T_p^\ell]\, ,
  \label{eq:turnstile}
  \end{equation}
where $\gamma_s^\ell = \{0,\, T_p^L\}$ defines the phase shift between
the leads ($\ell=L,R$) and $T_p^\ell = 30$ is the pulse length, the
same in the two leads.  The parameter $\delta$ controls the shape of
the pulse and is fixed at the value $\delta = 10$.  The time unit used
is $\hbar/t_s$.  The time dependent contact functions are graphed at
the bottom of Fig.\ \ref{fig:charge}.
The initial values are $\chi_{L,R} (0)=0$, i.\ e. the leads and the 
sample are initially disconnected.


\section{Results}\label{sec:results}

\begin{figure}[!ht]
    \centering
    \includegraphics{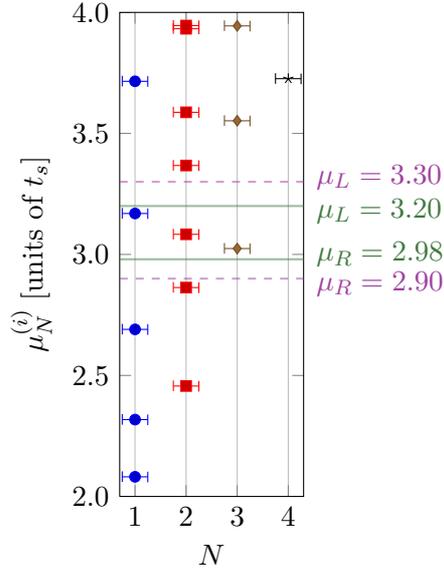}
    \caption{The $\mu$-diagram vs. the number of particles $N$ for $u_C
    = 0.5$.
The \textcolor{blue}{blue} points correspond to single-particle
states, the \textcolor{red}{red} points to two-particle states, the
\textcolor{brown!60!black}{brown} points to three-particle states,
and the only \textcolor{black}{black} point to four particles.
The bias window is $\Delta\mu=\mu_L-\mu_R$.  The small bias window is
$\Delta\mu=3.20-2.98=0.22$ (\textcolor{green!30!black!75}{green}).  The large one
is $\Delta\mu=3.30-2.90=0.40$ (\textcolor{violet!75}{violet}).  Both bias
windows include the ground state of three particles, but also excited
one and two-particle states.  So the expected number of electrons in
the steady state is slightly below three.
}
    \label{fig:mu-diag}
  \end{figure}

We chose the bias window $\Delta\mu=\mu_L-\mu_R$ to include the ground
state with $N=3$ electrons.  We performed transport calculations for
two bias windows: a larger one, with $\mu_L=3.30$ and $\mu_R=2.90$,
and a narrower one, with $\mu_L=3.20$ and $\mu_R=2.98$, as indicated
also in Fig.\ \ref{fig:mu-diag}.  In addition, we also collected the
results at vanishing bias $\mu_L=\mu_R=3.20$.

\begin{figure}[!ht]
    \centering
    \includegraphics[width=.45\textwidth]{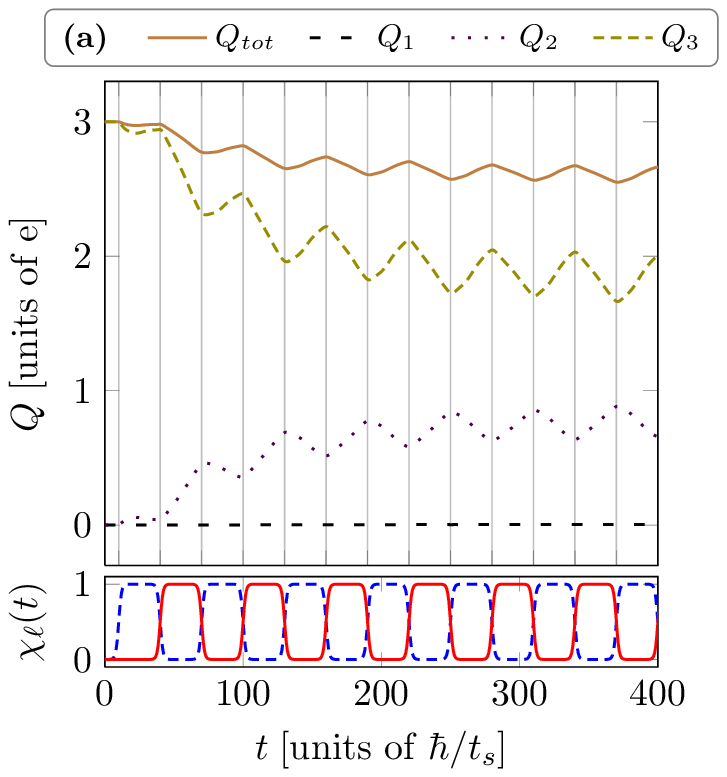}
    \includegraphics[width=.45\textwidth]{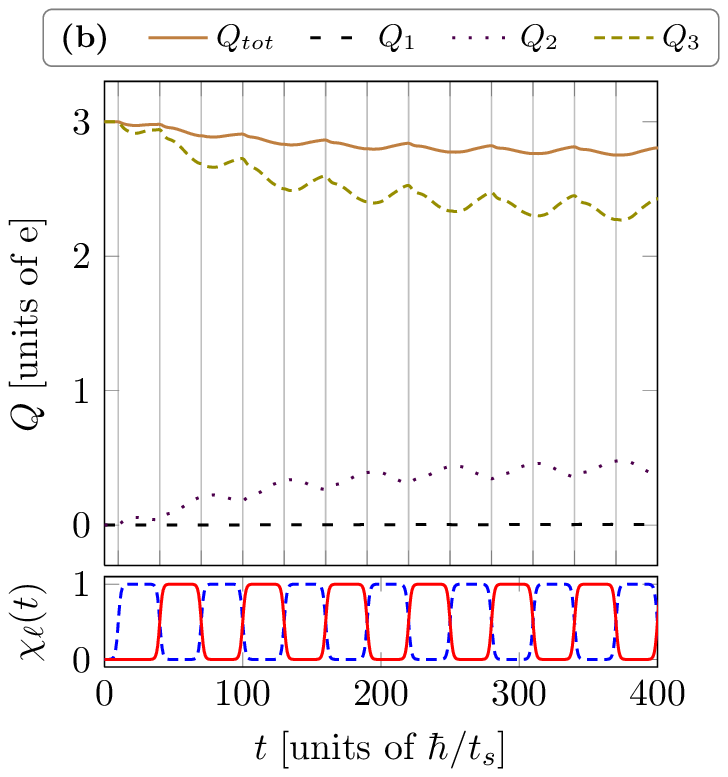}
    \caption{The time dependent charge for $\mu_L = 3.20$, $\mu_R = 2.98$. 
The contact functions $\chi_{L,R}$, given by Eq.\ \ref{eq:turnstile}, are shown at the bottom.
(a) Left lead at site 1 (\textcolor{blue}{blue}), right lead at site 10 (\textcolor{red}{red}).
(b) Left lead at site 1, right lead at site 3.
The total charge is shown by the continuous \textcolor{brown}{brown} line.  The dashed lines show the population
of the states with \textcolor{black}{1}, \textcolor{violet!60!black}{2}, and \textcolor{olive}{3} electrons
(indicated as $Q_1$, $Q_2$ and $Q_3$ on the top of the figures).
}
    \label{fig:charge}
  \end{figure}

We start the time-dependent calculations with $N=3$ electrons in the
sample, initially in the ground state.  We place the left lead ($L$) in
contact with site 1 of the sample (the left end) and the right lead ($R$)
in two different locations: first on site 10 (the right end) and then
on site 3.  In each case we turn on the turnstile, i.\ e.\ $\chi_{L,R}
(t)$ follow Eq.\ (\ref{eq:turnstile}), and calculate the time dependent
charge in the sample and the currents in the left and right leads.
In Fig.\ \ref{fig:charge} we show the evolution of the total charge in
the system for the two contact configurations mentioned above and the
narrower bias window.  The modulating signals impose charge oscillations,
which after some time become periodic.  Due to the choice of the bias
window the dominant populations correspond to the three particle ground
state and to a two-particle excited state.  Single particle states are
practically unpopulated and inactive in this case. We see that the
charge accumulated in the sample 'feels' the different placements of the
right lead. Indeed, when the drain lead is coupled to the site 3 the
oscillations of the total charge diminish compared to the case when
the drain is on site 10. One can say that part of the charge located
beyond the right contact, i.\ e.\ between sites 3 and 10, is somehow
frozen and does not contribute to transport. Note however than in both
configurations the populations of the two-particle and three-particle states
have opposite variations in time, i.\ e.\ the gain of one is partly compensated
by the loss of the other one.

\begin{figure}[!ht]
    \centering
    \includegraphics[width=.45\textwidth]{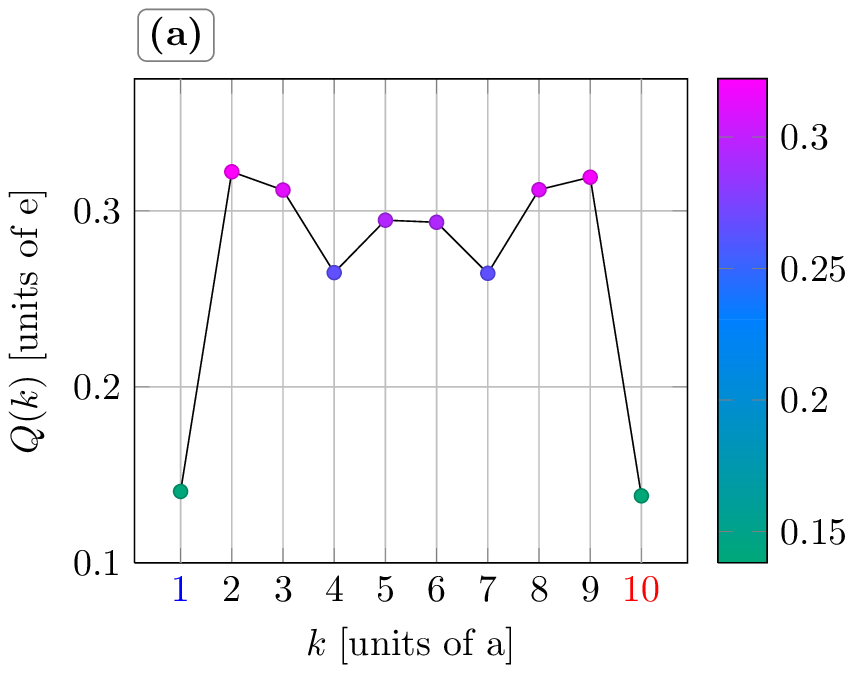}
    \includegraphics[width=.45\textwidth]{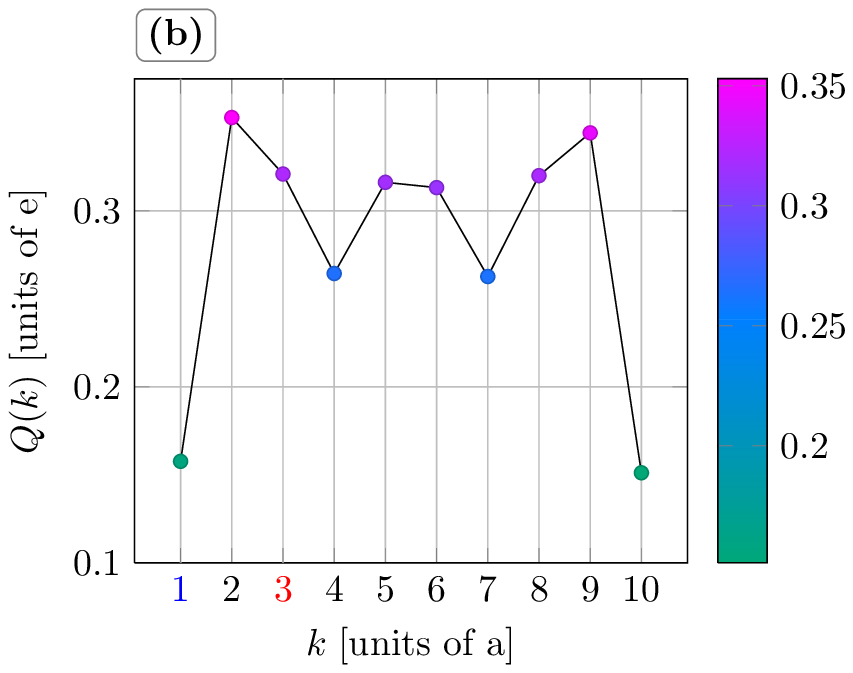}
    \caption{A snapshot of the charge distribution along the 10 sites,
for $\mu_L = 3.20$, $\mu_R = 2.98$. The time is $t=400$. The
position of the right contact is shown on the horizontal axis in \textcolor{red}{red}.
(a) Left 1 Right 10. (b) Left 1 Right 3.}
    \label{fig:charge_distrib}
  \end{figure}

The same behavior can be observed in the charge distribution along the
sample, which is shown in Fig.\ \ref{fig:charge_distrib} at a particular
time moment. The charge distribution is far from being homogeneous,
less charge being localized at the edges of the wire than in the center.
The charge distribution is also not symmetric along the sample, and at each
site it oscillates in time with frequencies related to the charging times of 
the active MES.  Consequently the currents in the leads also oscillate 
in time within each pumping period.

\begin{figure}[!ht]
    \centering
      \includegraphics[width=.45\textwidth]{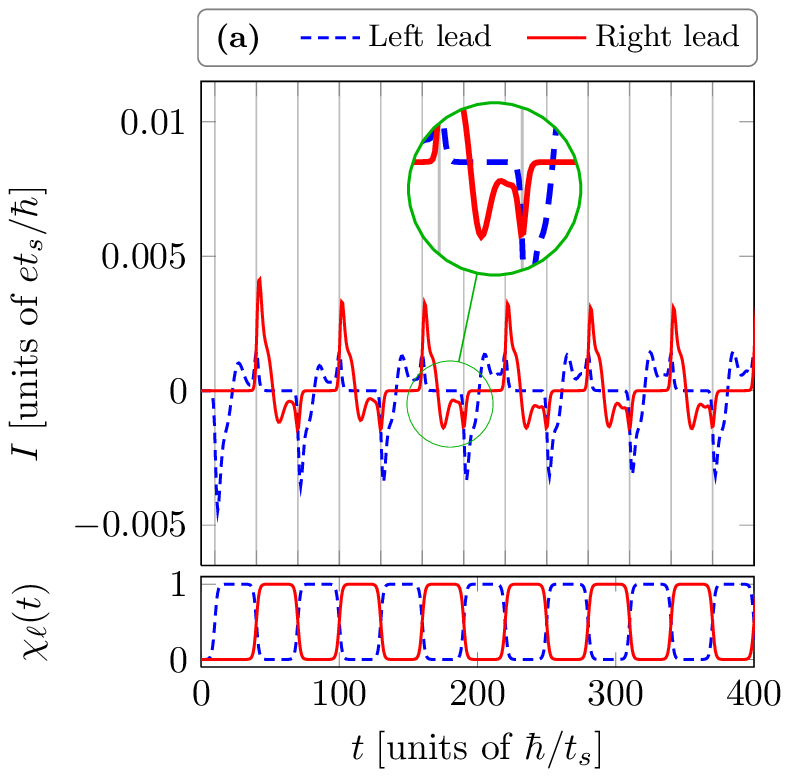}
      \includegraphics[width=.45\textwidth]{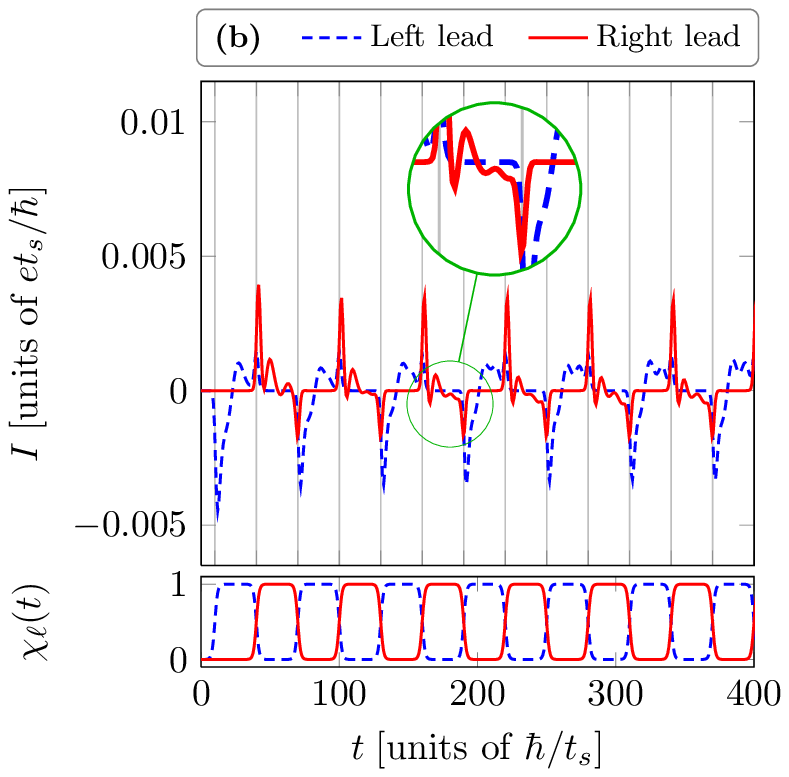}
      \caption{
The currents in the two leads at zero bias, $\mu_L = \mu_R = 3.20$:
(a) Left 1 Right 10. (b) Left 1 Right 3.
At $t=0$ the sample contains three electrons in the ground state.
}
    \label{fig:crt:zero_bias}
  \end{figure}

In the next figures we show the calculated currents in the two leads.  In Fig.\
\ref{fig:crt:zero_bias} we show the time dependent currents for zero bias.
The oscillations of the currents within each cycle average practically
to zero, so we can say that no net charge is transferred through the
sample.  The positive values of the currents correspond to the direction $L\to R$ and
the negative values to the opposite direction, $R\to L$.
One can notice some differences in the current oscillations
between the two placements of the right contacts, on site 10 and on site
3, respectively. More and sharper oscillations of the currents occur in
the later case, but still one can say that qualitatively the currents look
similar in the two cases.  

This situation changes totally with increasing the bias window.  In the
next figure, \ref {fig:crt:small_bias}, we show the currents for the small
bias window, i.\ e.\ $\mu_L=3.20$ and $\mu_R=2.98$.  Now the current
profiles for the two placements of the right contact are qualitatively
different.  With the right lead on site 10 we obtain positive currents
in both leads, describing charge flow in the direction imposed by the
the bias, i.\ e.\ from the left to the right of the system. Moreover,
the current in the left leads partially resembles the pulse shape.
But for the other placement, i.\ e.\ with the right contact at site 3,
we see much stronger oscillations, and even negative currents in both
leads for short times. These negative currents are actually {\em against}
the bias. In spite on their rather small amplitude one should note that
these negative currents do not vanish in the long time limit, that is
when the evolution of the system is periodic in time.  Another particular
feature of the asymmetric contact geometry is that it leads to pronounced
spikes in the left lead current.

\begin{figure}[!ht]
    \centering
      \includegraphics[width=.45\textwidth]{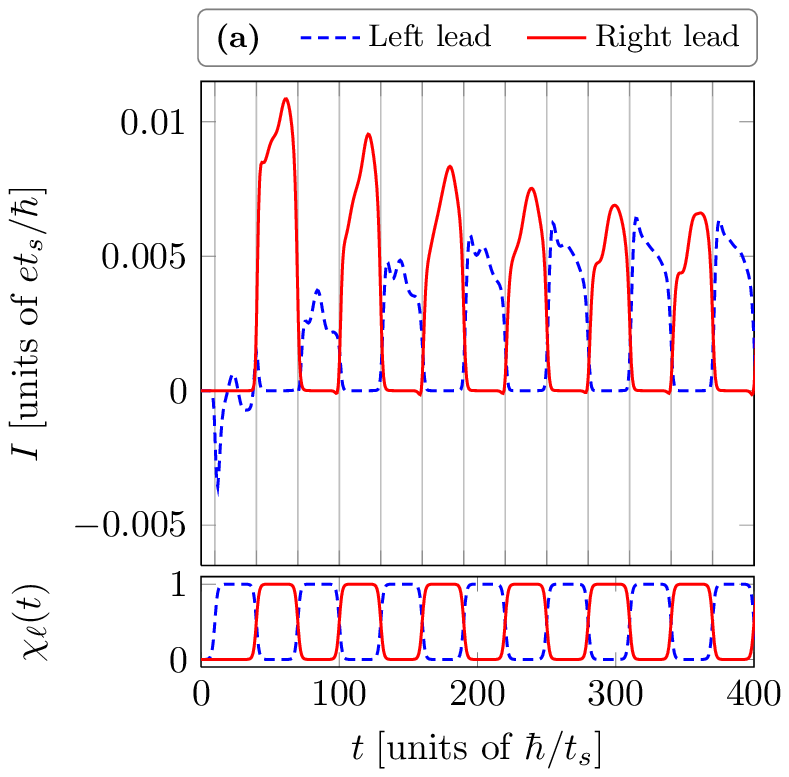}
      \includegraphics[width=.45\textwidth]{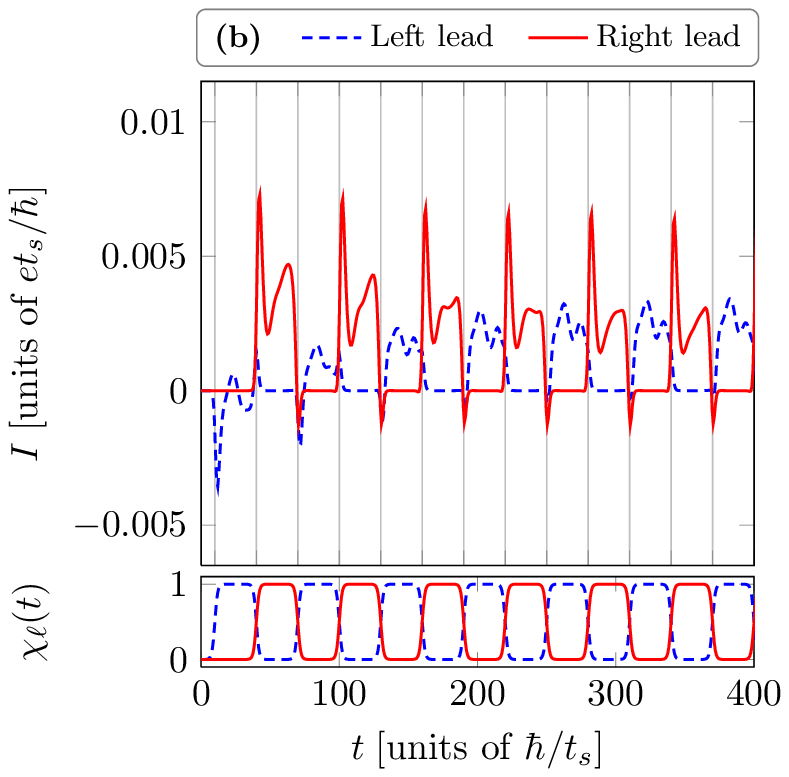}
      \caption{$\mu_L = 3.30$, $\mu_R = 2.90$. 
The currents in the two leads: 
 (a) Left 1 Right 10. (b) Left 1 Right 3.
The initial condition corresponds to three electrons in the ground state.
}
    \label{fig:crt:large_bias}
  \end{figure}

We then increase the bias window and obtain the results shown in Fig.\
\ref{fig:crt:large_bias}.  The current pulses in the contact placement
$L1-R10$ do not change qualitatively from the previous case of the smaller
bias window, but those for the placement $L1-R3$ do change: the negative
currents occur only in the right lead (the \textcolor{red}{red} solid
line), but not in the left lead (the \textcolor{blue}{blue} dashed line).
In this case some charge bounces back an forth between the sites 3 and
10, and enter or exit the sample through the $R$ lead, but it does not
reach the $L$ lead at site 1.  In other words the placement of the
right contact is qualitatively important for the current profiles for a 
finite bias window.  As we have checked, the negative currents survive
for a longer period of the pulses.

It is interesting to compare the current in one lead to the time
derivative of the charge, Figures\ \ref{fig:crt:small_bias} and
\ref{fig:charge}.  In the simplest interpretation, when the left
contact is open and the right contact is closed the charge in the sample
increases, and so the left current is positive and the right current
is zero.  In the next phase, when the left contact closes and the right
one opens, the charge decreases, and the right current is positive, the
left current being now zero.  The charging or discharging are actually
complex processes, because different states are occupied with different
time constants, and thus the charging is not constant in time, and the
currents have oscillations,  more or less following the time derivative
of the charge.  The fine structure of the currents is however more
complicated, also related to the charge oscillations on the particular
site where the contact is attached.

Finally, we want to comment on the Coulomb effects.
The Coulomb interaction which is included in the present calculations has
an important role in the charge distribution, through Coulomb
blocking and correlation effects, and so the currents are also affected. However, it
is quite difficult to compare the results without and with the Coulomb
interaction included. If Coulomb effects are neglected ($u_C=0$) the
whole energy spectrum changes and new states are present within the bias
window. Then the chemical potentials in the leads have to be shifted
accordingly in order to capture in the bias window states with similar
number of electrons as in the interacting case. This means one cannot
compare the two situations just by changing only one parameter.

  \begin{figure}[!ht]
    \centering
      \includegraphics[width=.45\textwidth]{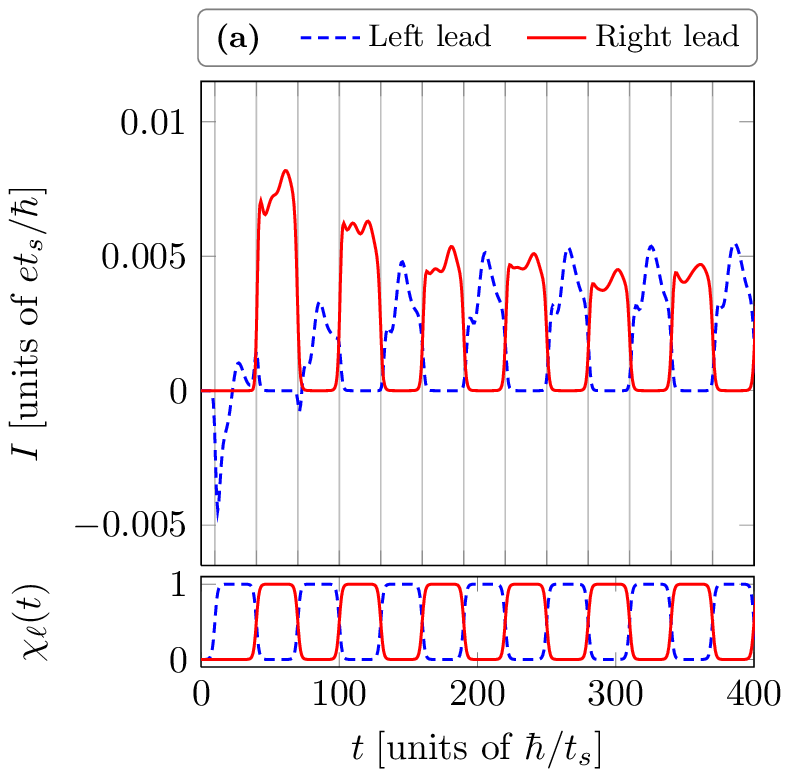}
      \includegraphics[width=.45\textwidth]{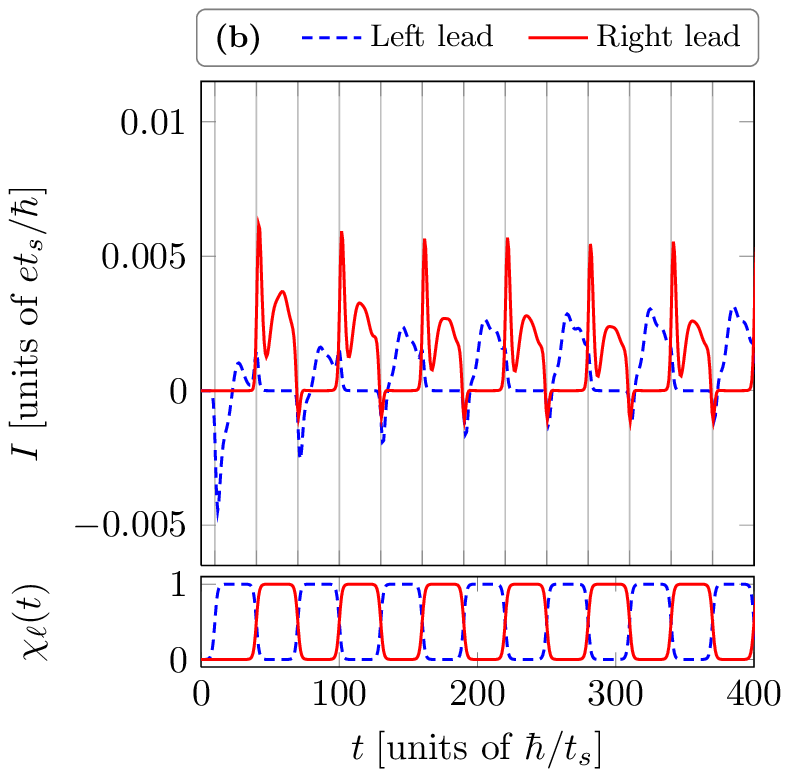}
      \caption{$\mu_L = 3.20$, $\mu_R = 2.98$.
 The currents in the two leads:
 (a) Left 1 Right 10. (b) Left 1 Right 3.
The initial condition corresponds to three electrons in the ground state.
}
    \label{fig:crt:small_bias}
  \end{figure}


\section{Concluding remarks}\label{sec:conclusion}

Using a lattice model we have simulated the time dependent transport
through a one-dimensional finite quantum wire attached to two
leads. Out-of-phase time-dependent signals are applied at the contact
regions, generating a turnstile device. The calculations are
performed by solving the generalized Master equation of the reduced
density operator. The latter acts in the Fock space of the many-body
states of the electrons in the sample, which are calculated via exact
diagonalization. We show that the location of the contacts with the leads
along the sample leaves clear fingerprints on the transient currents.
In particular, when changing the location of the drain lead the currents
may flow against the bias for a short time. Even though the currents
calculated in these examples are small and we are limited here only to
the qualitative effects, the observed counterflow may be seen in future
experiments. One can also expect that a suitable placement of the source
and drain leads along the sample would be a way to deliver modulated
output currents with a desired shape and period.

\ack
  This work was financially supported by the Icelandic Research Fund.
%



\begin{thebibliography}{10}
\expandafter\ifx\csname url\endcsname\relax
  \def\url#1{{\tt #1}}\fi
\expandafter\ifx\csname urlprefix\endcsname\relax\def\urlprefix{URL }\fi
\providecommand{\eprint}[2][]{\url{#2}}

\bibitem{Kurth1}
Kurth S, Stefanucci G, Almbladh C~O, Rubio A and Gross E~K~U 2005 {\em Phys.
  Rev. B\/} {\bf 72}(3) 035308

\bibitem{Stefanucci}
Stefanucci G and Almbladh C~O 2004 {\em Phys. Rev. B\/} {\bf 69}(19) 195318

\bibitem{Myohanen1}
My\"oh\"anen P, Stan A, Stefanucci G and van Leeuwen R 2009 {\em Phys. Rev.
  B\/} {\bf 80}(11) 115107

\bibitem{Myohanen2}
My\"oh\"anen P, Stan A, Stefanucci G and van Leeuwen R 2010 {\em J. Phys.:
  Conf. Series\/} {\bf 220} 012017

\bibitem{NJP}
Gainar C~M, Moldoveanu V, Manolescu A and Gudmundsson V 2011 {\em New. J.
  Phys.\/} {\bf 13} 013014

\bibitem{TSP}
Kouwenhoven L~P, Johnson A~T, van~der Vaart N~C, Harmans C~J~P~M and Foxon C~T
  1991 {\em Phys. Rev. Lett.\/} {\bf 67}(12) 1626--1629

\bibitem{Tarucha}
Fujisawa T, Austing D~G, Tokura Y, Hirayama Y and Tarucha S 2003 {\em J. Phys.
  Cond. Mat.\/} {\bf 15} R1395

\bibitem{Lai}
Lai W~T, Kuo D~M and Li P~W 2009 {\em Physica E\/} {\bf 41} 886 -- 889

\bibitem{Naser}
Naser B, Ferry D~K, Heeren J, Reno J~L and Bird J~P 2007 {\em Appl. Phys.
  Lett.\/} {\bf 90} 043103

\bibitem{Mizugaki}
Mizugaki Y 2003 {\em J. Appl. Phys\/} {\bf 94} 4480--4484

\bibitem{Ikeda}
Ikeda H and Tabe M 2006 {\em J. Appl. Phys\/} {\bf 99} 073705

\bibitem{PRBC}
Moldoveanu V, Manolescu A, Tang C~S and Gudmundsson V 2010 {\em Phys. Rev. B\/}
  {\bf 81}(15) 155442

\bibitem{PhysRevB.77.195416}
Timm C 2008 {\em Phys. Rev. B\/} {\bf 77}(19) 195416

\end{thebibliography}

\providecommand{\newblock}{}

\end{document}